\documentclass[conference]{IEEEtran}
\IEEEoverridecommandlockouts

\usepackage{cite}
\usepackage{amsmath,amssymb,amsfonts}
\usepackage{algorithmic}
\usepackage{graphicx}
\usepackage{textcomp}
\usepackage{xcolor}
\usepackage{subfigure}
\usepackage{booktabs}
\usepackage{multirow}
\usepackage{xcolor}
\usepackage{hyperref}
\hypersetup{
    colorlinks=true,
    urlcolor=magenta,
}

\def\BibTeX{{\rm B\kern-.05em{\sc i\kern-.025em b}\kern-.08em
    T\kern-.1667em\lower.7ex\hbox{E}\kern-.125emX}}
\begin{document}

\title{Training-Free Adaptive Quantization for \\ Variable Rate Image Coding for Machines }

\author{\IEEEauthorblockN{Yui Tatsumi}
\IEEEauthorblockA{\textit{Graduate School of FSE,} \\
\textit{Waseda University,}\\
Tokyo, Japan \\
yui.t@fuji.waseda.jp}
\and
\IEEEauthorblockN{Ziyue Zeng}
\IEEEauthorblockA{\textit{Graduate School of FSE,} \\
\textit{Waseda University,}\\
Tokyo, Japan \\
zengziyue@fuji.waseda.jp}
\and
\IEEEauthorblockN{Hiroshi Watanabe}
\IEEEauthorblockA{\textit{Graduate School of FSE,} \\
\textit{Waseda University,}\\
Tokyo, Japan \\
hiroshi.watanabe@waseda.jp}

}

\maketitle

\begin{abstract}
Image Coding for Machines (ICM) has become increasingly important with the rapid integration of computer vision technology into real-world applications. However, most neural network-based ICM frameworks operate at a fixed rate, thus requiring individual training for each target bitrate. This limitation may restrict their practical usage. 
Existing variable rate image compression approaches mitigate this issue but often rely on additional training, which increases computational costs and complicates deployment.
Moreover, variable rate control has not been thoroughly explored for ICM. 
To address these challenges, we propose a training-free framework for quantization strength control which enables flexible bitrate adjustment.
By exploiting the scale parameter predicted by the hyperprior network, the proposed method adaptively modulates quantization step sizes across both channel and spatial dimensions. 
This allows the model to preserve semantically important regions while coarsely quantizing less critical areas.
Our architectural design further enables continuous bitrate control through a single parameter.
Experimental results demonstrate the effectiveness of our proposed method, achieving up to 11.07\% BD-rate savings over the non-adaptive variable rate baseline. The code is available at \url{https://github.com/qwert-top/AQVR-ICM}.
\end{abstract}


\section{Introduction}
As research on computer vision advances, more image recognition models have become deeply integrated into real-world applications such as camera surveillance, smart agriculture, and autonomous driving. Since these applications continuously generate large amounts of data under limited network and storage resources, efficient compression optimized for machine analysis rather than human perception has become indispensable. 
Unlike conventional image compression designed for human eyes, Image Coding for Machines (ICM) aims to preserve only the information necessary for downstream machine vision tasks including classification, object detection, and segmentation \cite{ROI-first} - \cite{omni}.
Because machine vision tasks rely primarily on structural and semantic cues rather than perceptual fidelity, ICM models achieve reduced bitrates while preserving recognition accuracy. 

One of the major challenges in deploying Learned Image Compression (LIC)-based ICM models in practice lies in their limited ability to operate at variable bitrates. 
Traditional codecs such as JPEG and VVC \cite{VVC} can support a wide range of bitrates using a single model, simply by adjusting the quantization parameter.
In contrast, most existing LIC approaches \cite{balle} - \cite{LIC-last} require a distinct model for each bitrate, necessitating the training and storage of multiple parameter sets.
This limitation hinders their practical use, as real-world systems often require adaptive bitrate control depending on network bandwidth, device capability, or task requirements.

To address this, several variable rate LIC methods have been proposed \cite{variable RNN-first} - \cite{ECCV2024}. These approaches often require training with multiple rate-control knobs or conditional layers that enable a single network to operate across multiple bitrates. While effective, such training-based approaches require substantial computational costs, time, and complicated deployment pipelines. 
Furthermore, most of these methods are developed for human-oriented compression, and little attention has been paid to variable rate control specifically for ICM, despite its growing importance for machine-centric applications.
A straightforward, training-free alternative is to vary the quantization step size of a pretrained ICM model to control bitrate. 
However, this strategy injects noise uniformly across all latent features. While this may be tolerable for human viewing, it often leads to significant degradation in recognition performance, as task-relevant regions can be distorted.

In this paper, we propose a training-free variable rate framework for ICM. 
Our method enables flexible bitrate adjustment by dynamically controlling the quantization strength according to both channel-wise and spatial characteristics of the latent representation.
Consequently, it preserves semantically important regions while coarsely quantizing redundant areas.
The proposed design further allows continuous bitrate control with a single parameter, without requiring additional training or side information.
Owing to its simplicity and generality, our method can be readily integrated into existing ICM architectures for practical deployment.

\section{Related Work}
\begin{figure}
\centering
\subfigure[Average Scale of Each Channel]{%
\includegraphics[width=0.48\hsize]{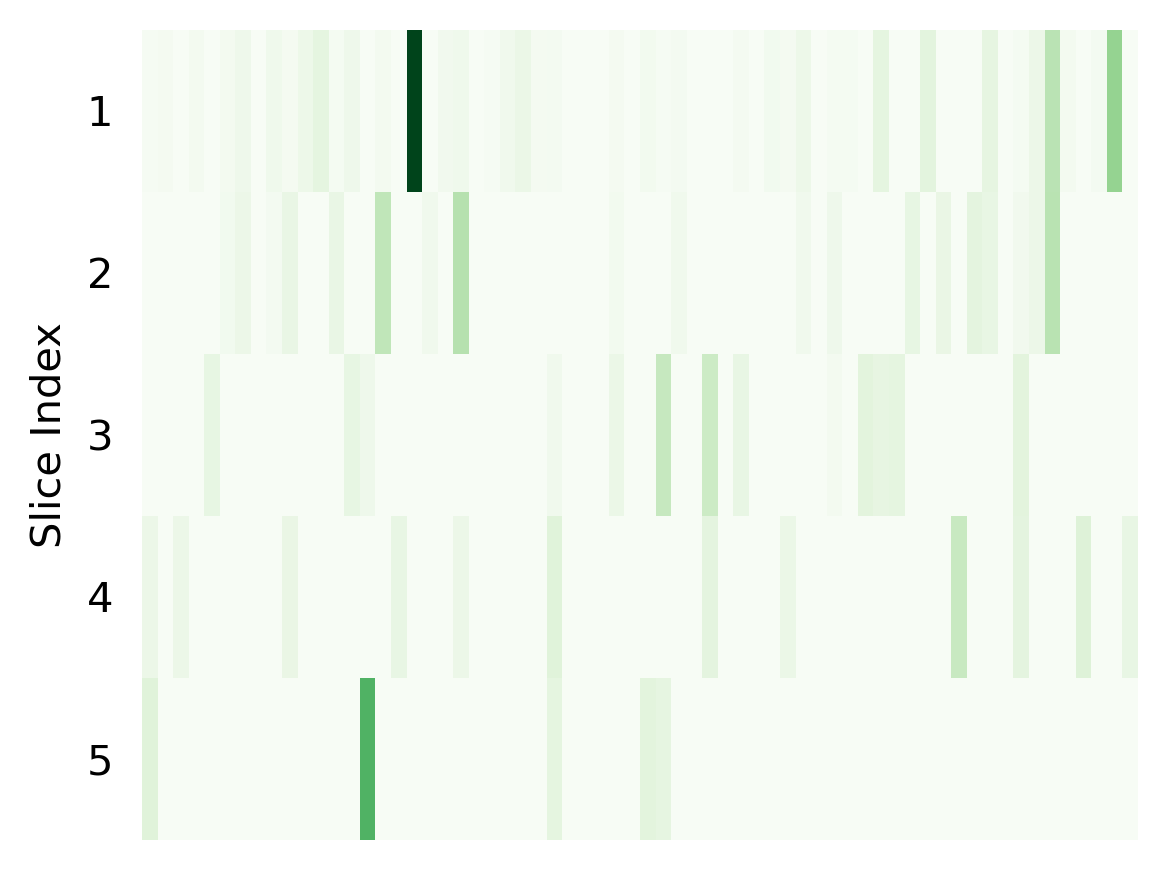}}%
\hspace{0.0025pt}
\subfigure[Average Scale of Each Slice]{%
\includegraphics[width=0.48\hsize]{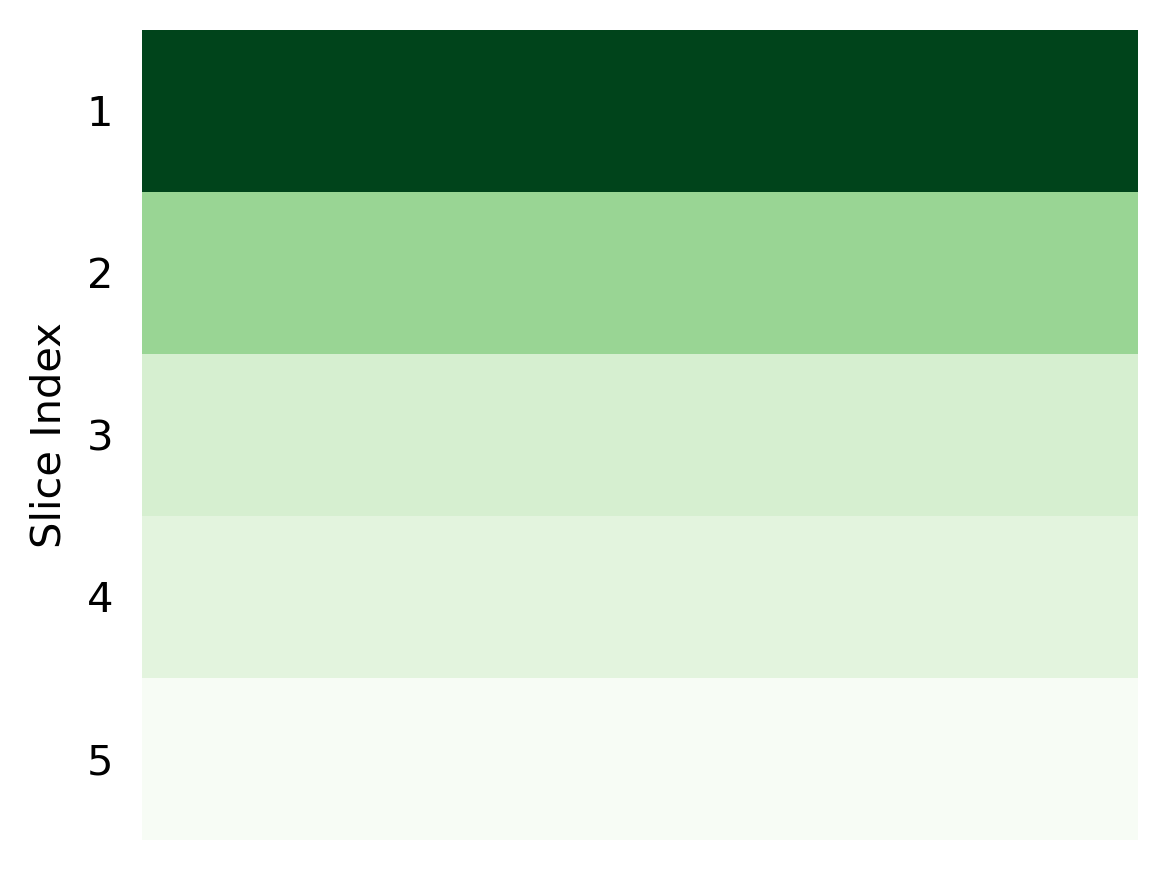}}%
\hspace{0.0025pt}
\caption{Visualization of average scale values of each (a) channel and (b) slice. Deeper colors represent larger values.}
\label{fig:entropy}
\end{figure}
\begin{figure}
\centering
\includegraphics[width=\hsize]{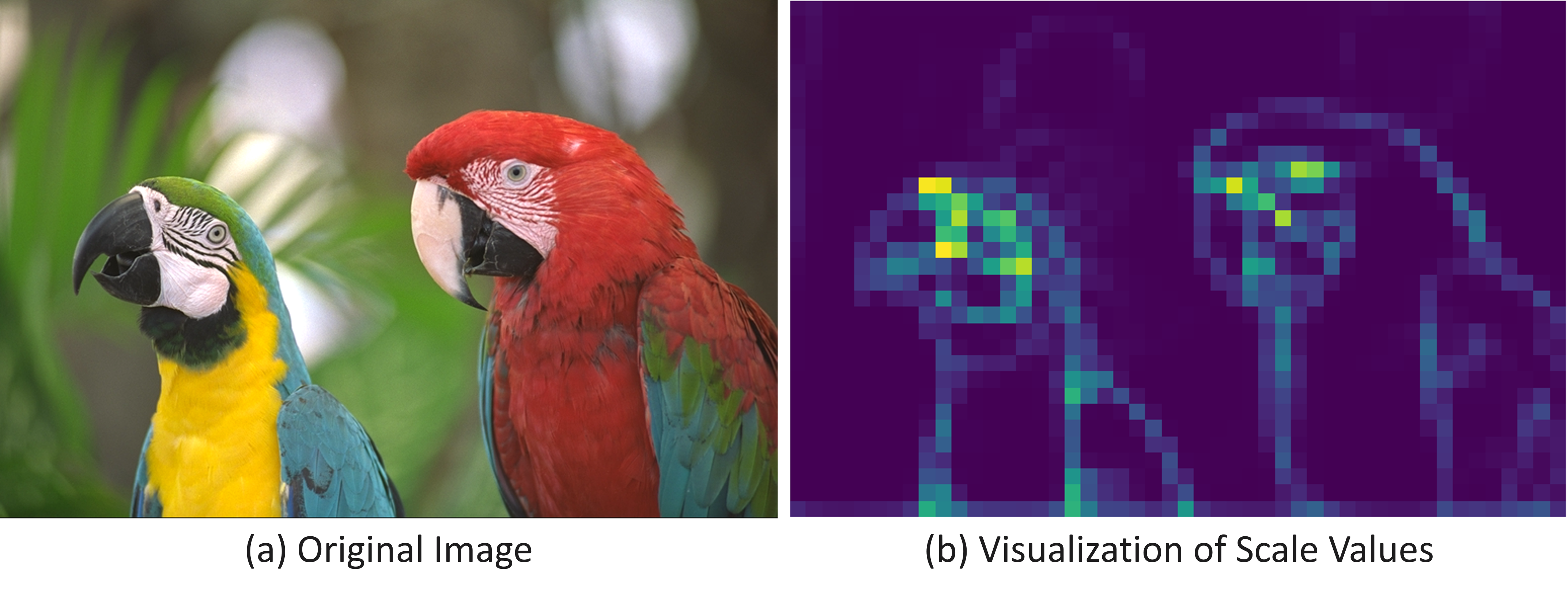}
\caption{Visualization of scale parameters in the first channel. Brighter colors indicate larger values. }
\label{fig:scale}
\end{figure}

\subsection{Learned Image Compression}
Representative LIC approaches are mainly based on variational autoencoders with a hyperprior network, as introduced by J. Ballé \textit{et al.} \cite{balle} and D. Minnen \textit{et al.} \cite{autoregressive}. 
Building on these foundations, the channel-wise autoregressive entropy model (Ch-ARM) \cite{ch-arm} divides the latent representation $\boldsymbol{y}$ along the channel dimension into $N$ slices $\{\boldsymbol{y}^{(1)}, \boldsymbol{y}^{(2)}, \dots, \boldsymbol{y}^{(N)}\}$, and the entropy parameters of each slice are predicted from the previously decoded ones.
This structure effectively captures inter-channel correlations, achieving higher compression efficiency and faster decoding.
Many subsequent LIC frameworks adopt Ch-ARM, such as TCM \cite{LIC-TCM}, which enhances the entropy model using a parameter-efficient Swin-Transformer–based attention module. 
D. He \textit{et al.} \cite{ELIC} further observe that in Ch-ARM, earlier slices tend to contain more critical information due to the sequential dependency. Based on this insight, they propose ELIC, which allocates higher capacity to earlier slices through an uneven grouping strategy, and achieve improved rate–distortion performance.
This observation motivates our channel-wise quantization control design described later in this paper.

Most LIC models share the following loss function, which jointly optimizes bitrate and reconstruction quality:
\begin{equation}
\mathcal{L} = \mathcal{R}(\boldsymbol{y}) + \mathcal{R}(\boldsymbol{z}) + \lambda \cdot mse(\boldsymbol{x}, \boldsymbol{\hat{x}}).
\label{eqn:lic-loss}
\end{equation}
In \eqref{eqn:lic-loss}, $\boldsymbol{y}$ and $\boldsymbol{z}$ are the outputs of the encoder and hyperprior-encoder of the LIC model, respectively. $\mathcal{R}(\cdot)$ denotes the estimated bitrate of each latent component. $\boldsymbol{x}$ represents the input image, and $\boldsymbol{\hat{x}}$ is the reconstructed image. $mse$ is the mean squared error function. 
The Lagrange multiplier $\lambda$ balances rate and distortion. Since it is fixed during training, multiple models with different $\lambda$ values must be trained to accommodate various rate–distortion trade-offs.

\subsection{Image Coding for Machines}
In recent years, more ICM methods have been proposed. 
ROI-based approaches \cite{ROI-first} - \cite{ROI-last} utilize region-of-interest (ROI) maps to allocate more bits to important regions, thus require prior analysis before compression. 
Task-loss-based methods \cite{task-first} - \cite{task-last} directly incorporate the performance of downstream recognition tasks into the training process. However, these methods must be retrained for each specific task and model, which limits their scalability in practical deployment. 
In contrast, region-learning-based approaches \cite{SA-ICM} - \cite{Delta-ICM} aim to achieve task-agnostic compression by learning to preserve spatial regions that are generally important for recognition. A notable example is SA-ICM \cite{SA-ICM}, which retains object boundaries while discarding other areas such as texture, as formulated by the following loss function: 
\begin{gather}
mask_x = canny(sam(\boldsymbol{x}, \alpha)), \label{eqn:mask} \\
\scalebox{0.95}{$
\mathcal{L} = \mathcal{R}(\boldsymbol{y}) + \mathcal{R}(\boldsymbol{z}) + 
\lambda \cdot mse(\boldsymbol{x} \odot mask_x, \boldsymbol{\hat{x}} \odot mask_x)
$}
  \label{eqn:saicm-loss}
\end{gather}
In \eqref{eqn:mask}, $sam$ and $canny$ denote region segmentation using the Segment Anything Model (SAM) \cite{sa} and Canny edge detection, respectively.
A constant value $\alpha$ denotes the confidence threshold for segmentation. SAM is used only during training.
SA-ICM is built upon LIC-TCM and thus employs the Ch-ARM.
Note that region-learning-based ICM requires fine-tuning of the downstream recognition model. This limitation, however, can be addressed by deploying fine-tuned recognition models on the cloud, whereas task-loss-based approaches demand separate task-specific codecs on edge devices. Additionally, SA-ICM can be extended to scalable image coding for humans and machines \cite{ICMH-FF} - \cite{MMSP2025}.

Delta-ICM \cite{Delta-ICM} further advances the ICM framework by redesigning the entropy model to adaptively switch between a Gaussian distribution for informative regions and a delta function for uninformative ones. This selective modeling allows less important areas to be decoded more coarsely, which results in greater bitrate savings without sacrificing recognition accuracy in downstream machine vision tasks. 
This finding motivates the spatially dependent quantization control introduced later in this paper.

\subsection{Variable Rate Image Compression}
While conventional LIC models are trained for a fixed rate, some studies have explored variable rate methods to accommodate diverse bandwidth and quality requirements.
Early works adopt Recurrent Neural Network (RNN)–based frameworks \cite{variable RNN-first}–\cite{variable RNN-last}, though they often require high computational cost due to their progressive coding.
Subsequent studies introduce conditional autoencoder–based approaches \cite{Y. Choi}–\cite{Z. Cui}, followed by quantization-based rate control methods \cite{QVRF}, \cite{F. Kamisli}.
For instance, QVRF \cite{QVRF} employs a quantization regulator to manage the overall quantization error of latent representations, enabling both discrete and continuous bitrate adjustment.
Another line of work explores mask-based selective coding. J. Lee \textit{et al.} \cite{Selective} achieve variable rate compression by selectively encoding essential latent representations using a learned 3D importance map generated from the hyperprior outputs.

Most of the previous approaches are designed for human-oriented compression, and only a few studies address variable bitrate control for ICM \cite{M. Song}, \cite{ECCV2024}.
M. Song \textit{et al.} \cite{M. Song} have proposed a spatially adaptive framework based on the Spatial Feature Transform, which enables variable rate compression conditioned on a pixel-wise quality map. This method can be extended to task-aware compression by optimizing the map for downstream recognition tasks.
Although existing variable rate methods for both LIC and ICM provide promising flexibility, they require training, which increases computational cost and complicates practical deployment. These challenges highlight the need for a training-free variable rate framework for ICM.

\section{Proposed Method}
\subsection{Preliminary Findings}
Since the scale parameter predicted by the hyperprior network determines bit allocation, we hypothesize that in ICM, it correlates with the feature importance required for recognition.
To investigate the channel-wise characteristics of the scale values, we conduct a preliminary experiment using an image from the Kodak dataset \cite{kodak}.
The image is compressed with SA-ICM, and the average scale values across channels and slices are visualized in Fig.~\ref{fig:entropy}. 
The results reveal that channels belonging to earlier slices tend to have larger scale values, which indicates the greater bit allocation, whereas those in later slices contain more redundant information.
This trend is consistent with the observations reported by D. He \textit{et al.} \cite{ELIC}, although their study utilized human-oriented LIC.

We also examine the spatial characteristics of the scale parameters, as illustrated in Fig.~\ref{fig:scale}.
Larger scale values align with object boundaries, while smaller values appear in other regions such as object interiors and backgrounds.
Given that SA-ICM is designed to preserve edge information effectively, these results indicate that the scale parameter reflects the region-wise importance for recognition.
This observation forms the key intuition behind our proposed method.
\begin{figure}
\centering
\includegraphics[width=\hsize]{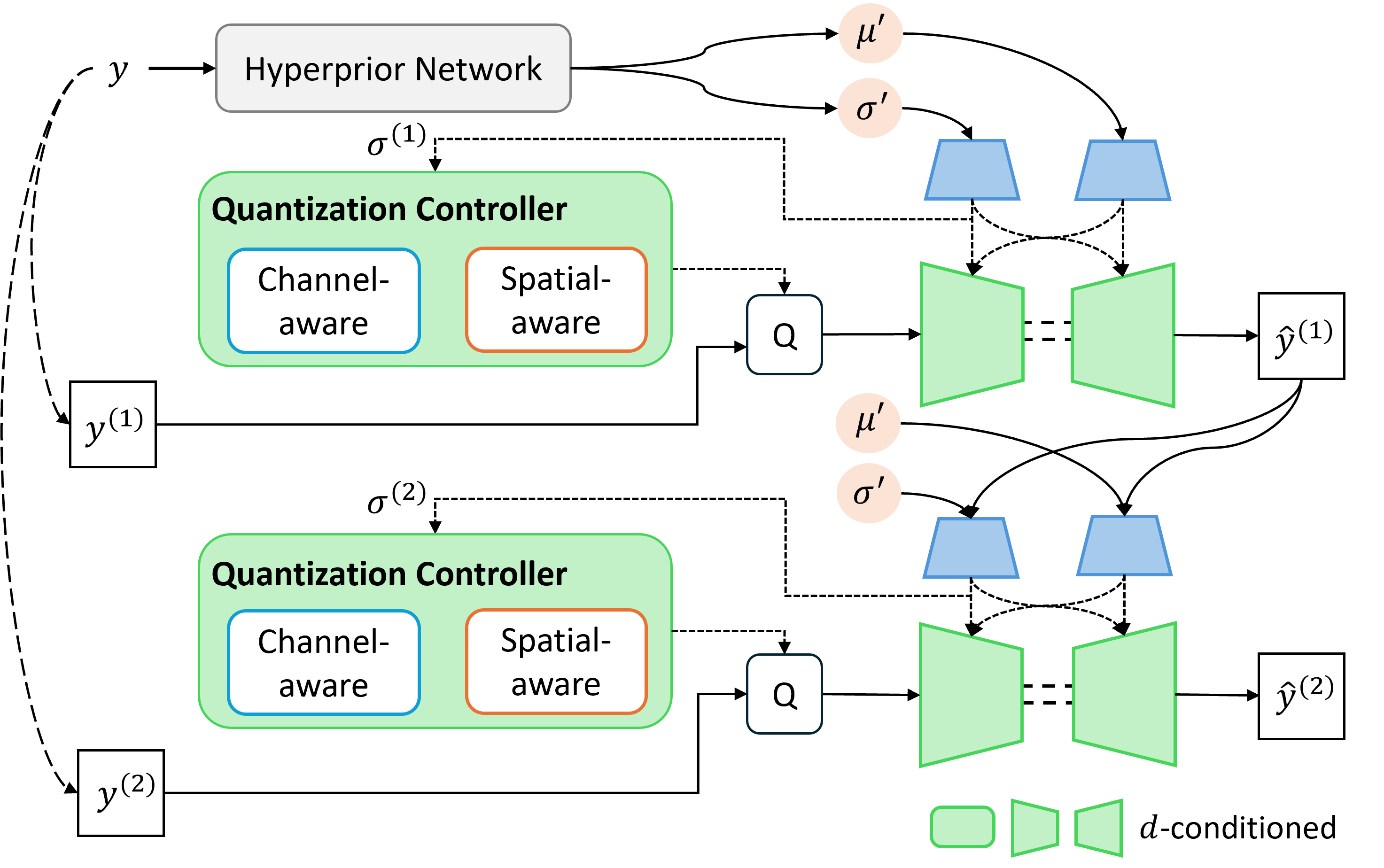}
\caption{Image coding process with the proposed method. The figure illustrates $N=2$ slices, whereas our experiments use $N=5$.}
\label{fig:framework}
\end{figure}


\begin{figure*}
\centering
\includegraphics[width=\hsize]{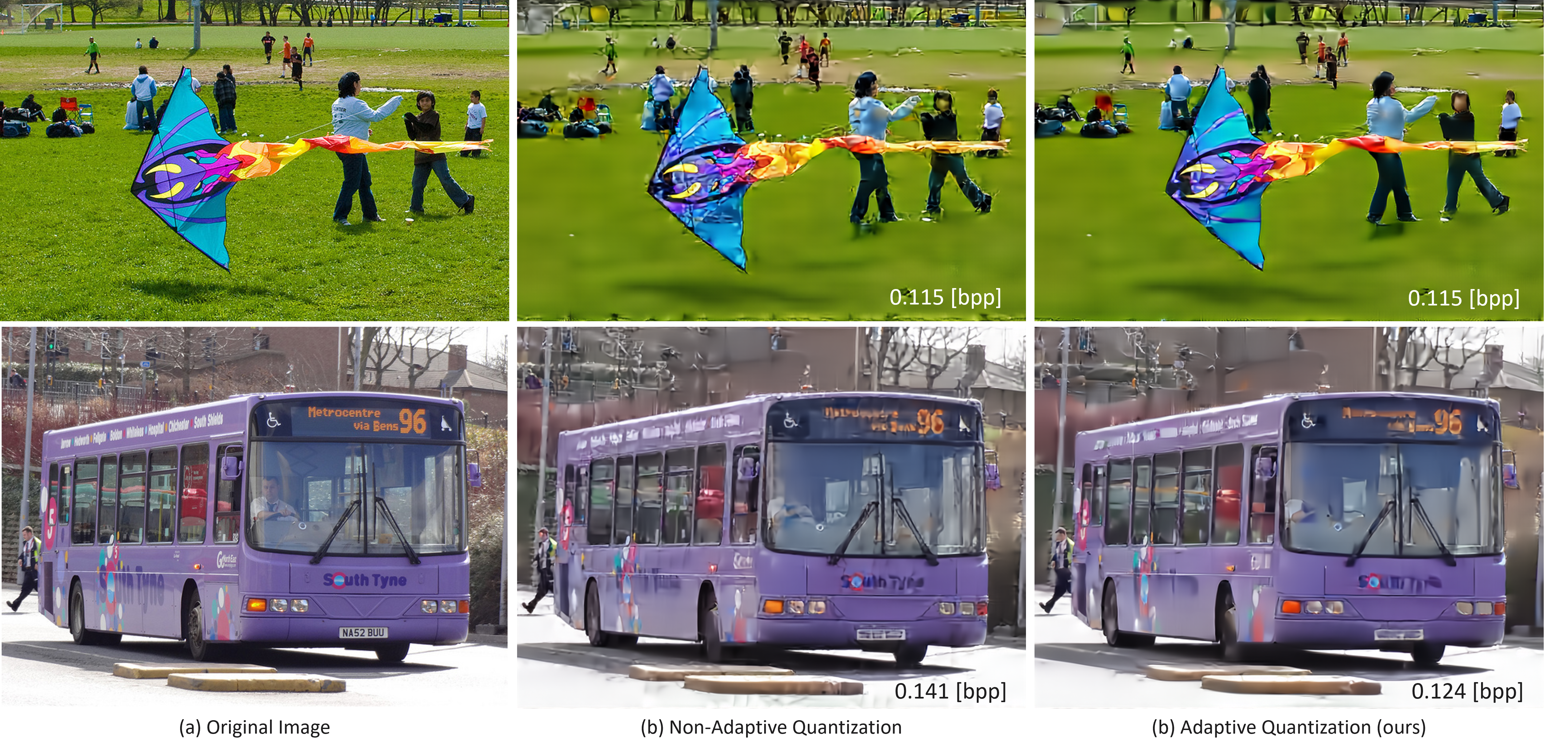}
\caption{Examples of reconstructed images with different quantization control methods. (a) Original image, (b) Non-adaptive quantization which uniformly scales the step size across all latents, and (c) Adaptive quantization step size control by the proposed method.}
\label{fig:outputs}
\end{figure*}

\subsection{Overview of Proposed Quantization Controller}
In this paper, we propose a training-free variable rate method for ICM. The compression process is illustrated in Fig. \ref{fig:framework}.
Since our preliminary findings reveal that the scale parameter serves as an indicator of feature importance for recognition, our approach adaptively adjusts the quantization strength according to this parameter to achieve variable bitrates without degrading recognition accuracy.
Specifically, the quantization controller first decides a channel-wise range of allowable quantization step sizes by utilizing scale characteristics. The actual step size for each latent pixel is then determined within this range based on spatial scale variations.
Note that the bitrate can be continuously controlled by a single parameter $d$, which facilitates deployment in real-world scenarios. The $d$ value is defined for $d > 0$; as $d$ increases, the transmitted bitrate decreases. When $d=1$, the bitrate matches that of the base model. 
While the quantization step size of the latent representation $\boldsymbol{y}$ is adaptively varied for rate control, that of the hyper latent $\boldsymbol{z}$ is fixed at 1 because coarse quantization of $\boldsymbol{z}$ would impair its role in predicting the variance of $\boldsymbol{y}$. Moreover, since the bitrate contribution of $\boldsymbol{z}$ is relatively small \cite{F. Kamisli}, its variation has negligible impact on overall rate control.

\subsection{Channel-Aware Quantization Step Size Control}
The preliminary experiments reveals that the scale parameter varies across channels. However, in Ch-ARM, the exact scales for all slices are not available simultaneously, making direct scheduling based on the true scales infeasible. 
Motivated by the empirical observation that earlier slices carry more critical information while later slices are increasingly redundant, we assign larger quantization step sizes to channels in later slices. 
Specifically, the user-specified parameter $d$ determines the quantization step size bounds. When $0< d < 1$, the minimum quantization step size for the $n$-th slice is linearly decreased, while when $d > 1$, the maximum size is linearly increased. The minimum and maximum quantization step size of the $n$-th slice, $\Delta_{\min} ^{(n)}$ and $\Delta_{\max} ^ {(n)}$, are defined as:
\begin{equation}
\Delta_{\min} ^{(n)} =
\begin{cases}
d + \dfrac{n-1}{N}\,(1-d), & 0<d<1,\\[6pt]
1, & d=1,\\[6pt]
1, & d>1,
\end{cases}
\end{equation}
\begin{equation}
\Delta_{\max} ^ {(n)} =
\begin{cases}
1, & 0<d<1,\\[6pt]
1, & d=1,\\[6pt]
1 + \dfrac{n}{N}\,(d-1), & d>1,
\end{cases}
\end{equation}
where $N$ denotes the number of slices and $n = 1, 2, \dots, N$ represents the slice index.
This design ensures that earlier slices retain finer details, while later slices are quantized more coarsely, preserving the slice-wise importance.

\subsection{Spatial-Aware Quantization Step Size Control}
Within each channel, the quantization step size is further adapted per spatial position according to the scale parameter predicted by the hyperprior network.
Given the scale tensor $\boldsymbol{\sigma}$ and the slice-wise bounds $\Delta_{\min} ^{(n)}$ and $\Delta_{\max} ^ {(n)}$ determined above,
the adaptive quantization step size
$\Delta_{c,h,w}$ is computed as
\begin{gather}
\sigma_{\max} ^ {(c)} 
= \max_{h,w}\, \sigma_{c,h,w}, \\
\sigma_{\min} ^{(c)} 
= \min_{h,w}\, \sigma_{c,h,w}, \\
\Delta_{c,h,w}
= \Delta_{\max} ^ {(n)}
- \frac{(\sigma_{c,h,w} - \sigma_{\min} ^ {(c)})
(\Delta_{\max} ^ {(n)} - \Delta_{\min} ^ {(n)})}
{\sigma_{\max}^{(c)} - \sigma_{\min}^{(c)} + \epsilon},
\label{eqn:adaptive_delta_all}
\end{gather}
where $\sigma_{min}^{(c)}$ and $\sigma_{max}^{(c)}$ denote
the minimum and maximum scale values for each channel, and $\epsilon$ is a small constant.
This linear mapping assigns smaller quantization steps to regions with higher scale values and larger steps to those with lower scale, enabling spatially adaptive quantization.
Mapping functions other than linear are evaluated in the ablation study. 

\subsection{Adaptive Quantization in the Compression Process}
During compression, the main encoder extracts the latent feature $\boldsymbol{y}$ from an input image, which is then quantized. The quantization is usually performed by the following equation:
\begin{equation}
\boldsymbol{\hat{y}} = round(\boldsymbol{y}-\boldsymbol{\mu}) + \boldsymbol{\mu},
\end{equation}
where $\boldsymbol{\mu}$ is the predicted mean parameter.
In our proposed method, each latent slice is quantized using the adaptive
$\Delta$ tensor determined by the quantization controller:
\begin{equation}
\boldsymbol{\hat{y}} = round(\frac{\boldsymbol{y}-\boldsymbol{\mu}}{\boldsymbol{\Delta}}) \times \boldsymbol{\Delta}+ \boldsymbol{\mu}.
\label{eq:quant}
\end{equation}

The quantized feature is then compressed using an entropy model, which predicts the mean and scale parameters of the encoding distribution.
In standard LIC models which assume a Gaussian distribution and a unit quantization step size, the entropy model is formulated as follows:
\begin{align}
    p_{\boldsymbol{\hat{y}}|\boldsymbol{\hat{z}}}(\boldsymbol{\hat{y}}|\boldsymbol{\hat{z}}) &= \prod_{i}p_{\boldsymbol{\hat{y}}|\boldsymbol{\hat{z}}}(\hat{y_i}|\boldsymbol{\hat{z}}), \\
    p_{\boldsymbol{\hat{y}}|\boldsymbol{\hat{z}}}(\hat{y_i}|\boldsymbol{\hat{z}})&= \mathcal{N}(\mu_i, {\sigma_i}^2)*\mathcal{U}\left(-\frac{1}{2}, \frac{1}{2}\right)(\hat{y_i}), \\
    &= \int_{\hat{y_i}-1/2}^{\hat{y_i}+1/2} \mathcal{N}(t |\mu_i, {\sigma_i}^2)dt.
    \label{eqn:unit_quantize}
\end{align}
However, in our proposed method, the quantization step size is adaptively varied.
To correctly reflect this operation in the probability formulation, we modify the Gaussian model as expressed in the following equation:
\begin{align}
    p_{\boldsymbol{\hat{y}}|\boldsymbol{\hat{z}}}(\hat{y_i}|\boldsymbol{\hat{z}}, \Delta_i)&= \int_{\hat{y_i}-\Delta_i/2}^{\hat{y_i}+\Delta_i/2} \mathcal{N} (t |\mu_i, \sigma_i^2) dt.
\end{align}
Note that this modification is only applied during the inference. The entropy model itself and its parameters remain unchanged. 
\begin{figure*}
\centering
\subfigure[Object Detection by YOLOv5]{%
\includegraphics[width=0.32\hsize]{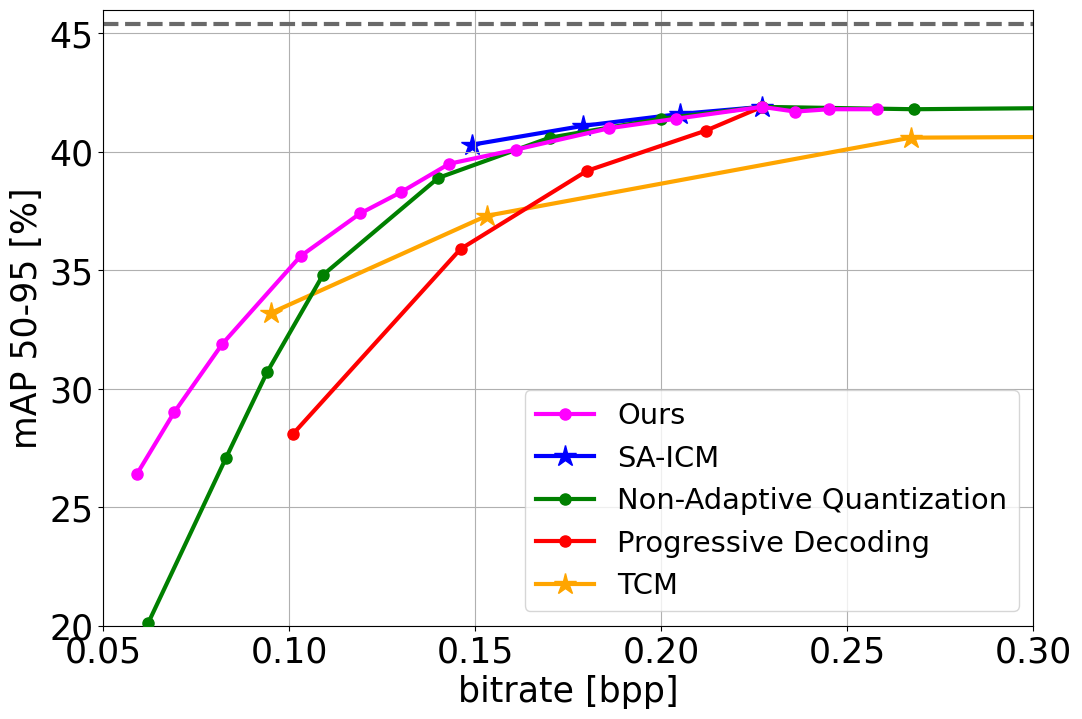}}%
\hspace{0.0025pt}
\subfigure[Object Detection by Mask R-CNN]{%
\includegraphics[width=0.32\hsize]{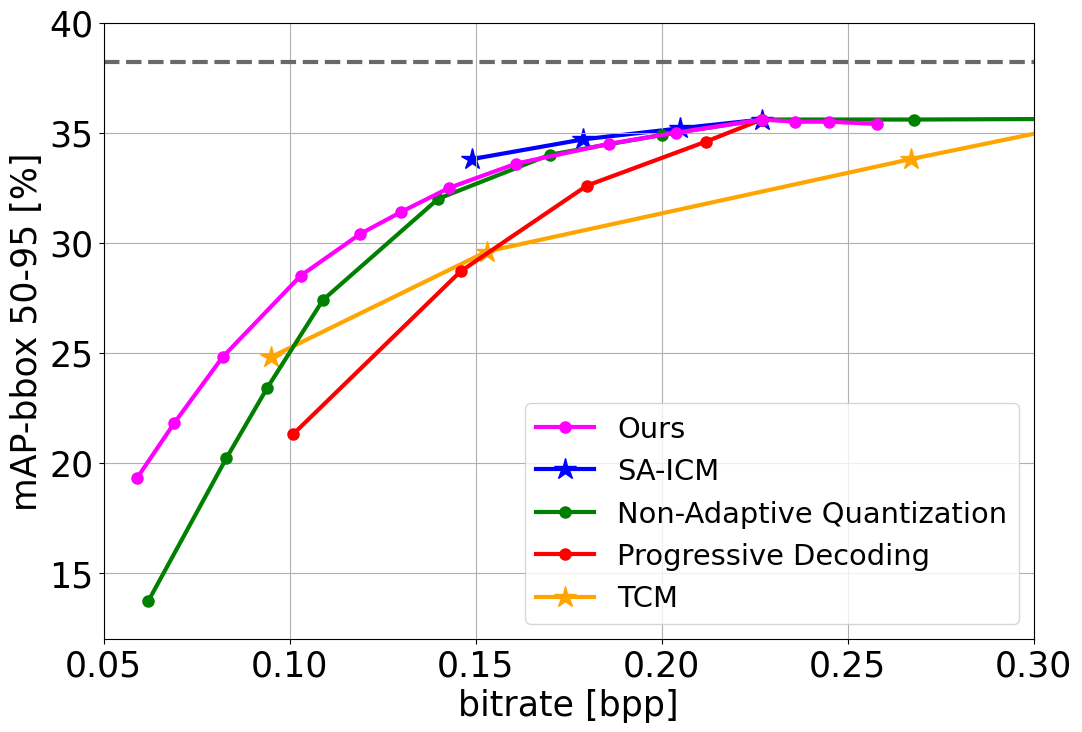}}%
\hspace{0.0025pt}
\subfigure[Instance Segmentation by Mask R-CNN]{%
\includegraphics[width=0.32\hsize]{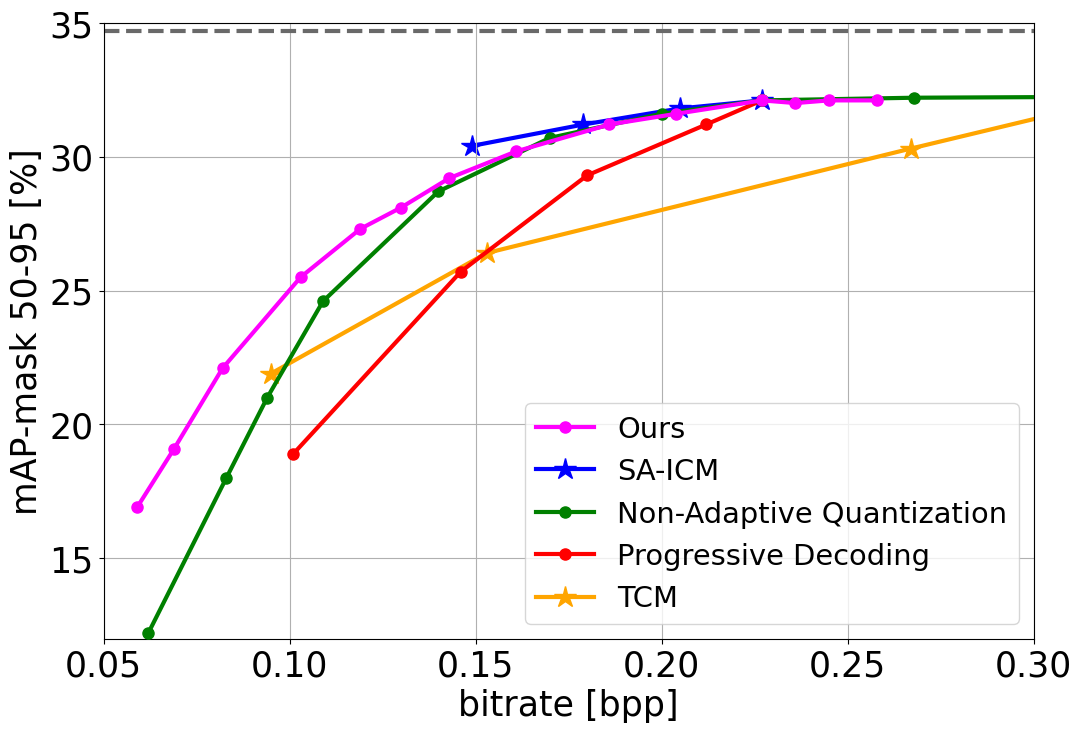}}%
\hspace{0.0025pt}
\caption{Image compression performance for different recognition tasks. (a) Object detection by YOLOv5, (b) Object detection by Mask R-CNN, and (c) Instance segmentation by Mask R-CNN.}
\label{fig:rd-curve}
\end{figure*}

\begin{figure}
\centering
\includegraphics[width=\hsize]{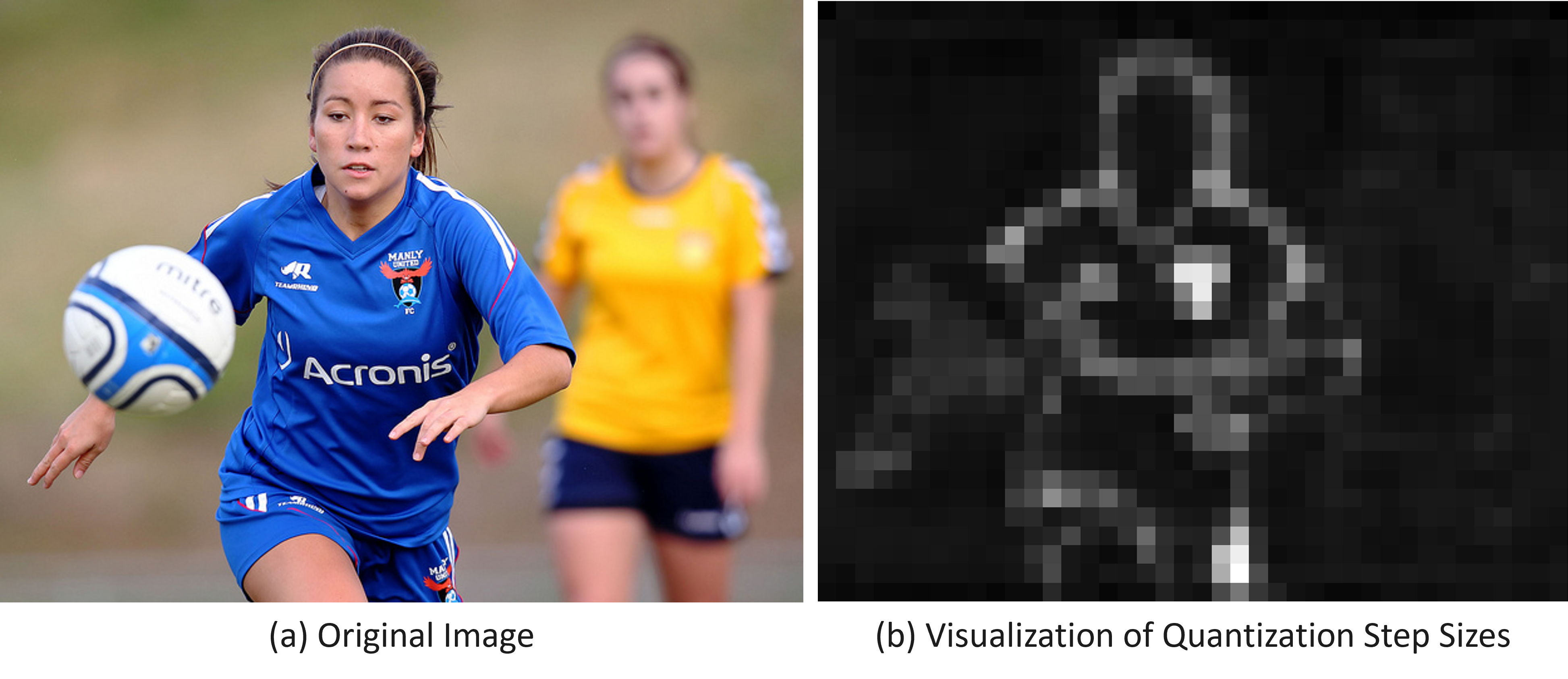}
\caption{Visualization of quantization step sizes in the first channel determined by the proposed method when $d=8$. White colors indicate smaller steps.}
\label{fig:delta}
\end{figure}
\section{Experiment}
\subsection{Experimental Settings}
We evaluate our proposed method on multiple machine vision tasks using SA-ICM as a base model, though the proposed quantization control scheme can be applied to other ICM frameworks with Ch-ARM structure. SA-ICM is pre-trained on 118,287 images from the COCO-train dataset \cite{coco}, with a fixed confidence threshold $\alpha = 0.78$, $N=5$ latent slices, and $\lambda = 0.05$ in the loss function represented in \eqref{eqn:saicm-loss}. This base model achieves 0.227 [bpp] on average. By varying the quantization step size with parameter $d$, we achieve variable bitrates. Evaluation is performed on 5,000 images from the COCO-val dataset.
We compare our method against 1) Fixed-rate SA-ICM trained separately with $\lambda = \{0.02, 0.03, 0.04, 0.05\}$,  2) Non-Adaptive Quantization, which utilizes a single quantization step size globally without adaptive control,
3) Progressive decoding introduced by D. Minnen \textit{et al.} \cite{ch-arm}, where only the first $n$ slices are transmitted, and the rest are substituted with hyperprior-predicted mean values to achieve discrete bitrate control by varying $n$, 
and 4) Fixed-rate LIC-TCM, a codec for human perception. The second and third methods are training-free variable rate approaches. 

We assess the recognition accuracy of compressed images using YOLOv5 \cite{Yolov5} for object detection and Mask R-CNN \cite{mask r-cnn} for both object detection and instance segmentation. These models are selected for comparison with prior work, though our method can be applied to any machine vision model. 
All recognition models are fine-tuned on COCO-train images decoded by SA-ICM with $\lambda = 0.05$ for the proposed and first three comparative methods, and by TCM with $\lambda = 0.05$ for the last one. 

\subsection{Experimental Results}
Example outputs are shown in Fig. \ref{fig:outputs}. Compared to the Non-Adaptive Quantization approach, our proposed method preserves clearer object boundaries crucial for recognition tasks, while achieving equal or even lower bitrates.
The Rate-mAP curves of the proposed and comparative methods are illustrated in Fig. \ref{fig:rd-curve}. The gray dotted lines represent the recognition accuracy on uncompressed images.
Marker shapes denote method categories: fixed-rate models are plotted with stars, whereas training-free variable rate methods are plotted with circles.
It is shown that the variable rate model has almost the same recognition accuracy even around 0.15 [bpp], which is the lowest bitrate for pretrained SA-ICM model. Our proposed method outperforms the baseline method especially at lower bitrates. Fig. \ref{fig:delta} visualizes the spatial distribution of quantization step size assigned by the proposed method when $d=8$. It clearly shows that the smaller steps are selected especially in object boundaries, which are important for recognition.

Moreover, we measure encoding and decoding times on COCO-val dataset at $d=8$, averaging over five runs. The proposed quantization controller introduces only minor differences compared to the fixed-rate SA-ICM model, 
with the encoding time increased by approximately 2 [ms] $(\approx1.5\%)$ and the decoding time by 2 [ms] $(\approx1.3\%)$.
These results confirm that our adaptive quantization method adds negligible computational overhead.
\begin{table}[t]
  \centering
  \caption{BD-rate-mAP comparison of different spatial-aware mappings}
\begin{tabular}{cccc}
  \toprule
  \multirow{2}{*}{Method} &
  \multicolumn{2}{c}{Detection} &
  \multicolumn{1}{c}{Segmentation} \\
  \cmidrule(lr){2-3} \cmidrule(lr){4-4}
    & YOLOv5 & Mask R-CNN & Mask R-CNN \\
  \midrule
  Non-Adaptive Quantization       & 0.00 & 0.00 & 0.00 \\
  Sigmoid ($k=5$) & -10.84 &-10.61 & \textbf{-11.41}\\
  Sigmoid ($k=10$) & -8.82 & -8.53 & -7.95 \\
  Linear (ours)        & \textbf{-11.07} & \textbf{-10.88} & -10.37 \\
  \bottomrule
\end{tabular}
\label{tab:ablation2}
\end{table}
\subsection{Ablation Study}
We conduct an ablation study to investigate different scheduling strategies for the spatial-aware quantization control.
Table \ref{tab:ablation2} shows mAP-based BD-rates for the proposed linear mapping and sigmoid-based mappings, measured relative to the Non-Adaptive Quantization approach. The sigmoid mapping is defined as follows:
\begin{gather}
\sigma_{\text{norm}}^{(c)} = \frac{\sigma_{c, h, w} - \sigma_{\min}^{(c)}}{\sigma_{\max}^{(c)} - \sigma_{\min}^{(c)} + \epsilon}, \\
\hspace{1pt}
\scalebox{0.8}{$
\Delta_{c, h, w} = \Delta_{\min}^{(n)} + \text{sigmoid}\left( -k \left( \sigma_{\text{norm}}^{(c)} - 0.5 \right) \right) (\Delta_{\max}^{(n)} - \Delta_{\min}^{(n)}).
$}
\label{eqn:sigmoid}
\end{gather}
In \eqref{eqn:sigmoid}, $\text{sigmoid}(x)$ denotes the sigmoid function defined as $1/(1+\exp(-x))$, and the other variables are the same as in \eqref{eqn:adaptive_delta_all}.
The linear mapping achieves superior or comparable performance across recognition tasks, indicating its effectiveness as a simple yet robust scheduling strategy. 

\section{Conclusion}
In this paper, we propose a training-free quantization controller for variable rate ICM that adaptively adjusts quantization step size based on channel-wise and spatial characteristics. By assigning larger quantization steps to later channels and redundant regions according to the scale parameter predicted by the hyperprior network, our method effectively reduces bitrate without training.
Additionally, the bitrate can be continuously controlled by a single parameter.
Experimental results demonstrate that our proposed method outperforms a variable rate baseline with non-adaptive quantization. 
Future work will further extend this framework to scalable image coding for humans and machines.

\section*{Acknowledgment}
The results of this research were obtained from the commissioned research (JPJ012368C05101) by National Institute of Information and Communications Technology (NICT), Japan.

\end{document}